\documentclass[longauth]{aa} 
\usepackage{graphicx}
\usepackage{lscape}
\begin{document}
  \title{Multi-wavelength
	Observations of H\,2356$-$309}

\author{HESS Collaboration
 \and A.~Abramowski \inst{4}
 \and F.~Acero \inst{15}
 \and F. Aharonian\inst{1,13}
 \and A.G.~Akhperjanian \inst{2}
 \and G.~Anton \inst{16}
 \and U.~Barres de Almeida \inst{8} \thanks{supported by CAPES Foundation, Ministry of Education of Brazil}
 \and A.R.~Bazer-Bachi \inst{3}
 \and Y.~Becherini \inst{12}
 \and B.~Behera \inst{14}
 \and W.~Benbow \inst{1}  \thanks{now at Harvard-Smithsonian Center for Astrophysics, Cambridge, MA, USA}
% ...
}

\offprints{wbenbow@cfa.harvard.edu or catherine.boisson@obspm.fr}

\institute{
Max-Planck-Institut f\"ur Kernphysik, P.O. Box 103980, D 69029
Heidelberg, Germany
\and
 Yerevan Physics Institute, 2 Alikhanian Brothers St., 375036 Yerevan,
Armenia
\and
Centre d'Etude Spatiale des Rayonnements, CNRS/UPS, 9 av. du Colonel Roche, BP
4346, F-31029 Toulouse Cedex 4, France
\and
Universit\"at Hamburg, Institut f\"ur Experimentalphysik, Luruper Chaussee
149, D 22761 Hamburg, Germany
\and
Institut f\"ur Physik, Humboldt-Universit\"at zu Berlin, Newtonstr. 15,
D 12489 Berlin, Germany
\and
LUTH, Observatoire de Paris, CNRS, Universit\'e Paris Diderot, 5 Place Jules Janssen, 92190 Meudon, 
France
\and
CEA Saclay, DSM/IRFU, F-91191 Gif-Sur-Yvette Cedex, France
\and
University of Durham, Department of Physics, South Road, Durham DH1 3LE,
U.K.
\and
Unit for Space Physics, North-West University, Potchefstroom 2520,
    South Africa
\and
Laboratoire Leprince-Ringuet, Ecole Polytechnique, CNRS/IN2P3,
 F-91128 Palaiseau, France
\and 
Laboratoire d'Annecy-le-Vieux de Physique des Particules,
Universit\'{e} de Savoie, CNRS/IN2P3, F-74941 Annecy-le-Vieux,
France
\and
Astroparticule et Cosmologie (APC), CNRS, Universit\'{e} Paris 7 Denis Diderot,
10, rue Alice Domon et L\'{e}onie Duquet, F-75205 Paris Cedex 13, France
\thanks{UMR 7164 (CNRS, Universit\'e Paris VII, CEA, Observatoire de Paris)}
\and
Dublin Institute for Advanced Studies, 5 Merrion Square, Dublin 2,
Ireland
\and
Landessternwarte, Universit\"at Heidelberg, K\"onigstuhl, D 69117 Heidelberg, Germany
\and
Laboratoire de Physique Th\'eorique et Astroparticules, 
Universit\'e Montpellier 2, CNRS/IN2P3, CC 70, Place Eug\`ene Bataillon, F-34095
Montpellier Cedex 5, France
\and
Universit\"at Erlangen-N\"urnberg, Physikalisches Institut, Erwin-Rommel-Str. 1,
D 91058 Erlangen, Germany
% ...
}

  \date{Received; accepted}


 \abstract
  {}
  {The properties of the broad-band emission from the high-frequency
peaked BL\,Lac H\,2356$-$309  ($z=0.165$) are investigated.}
  {Very High Energy (VHE; E $>$ 100 GeV) observations of H\,2356$-$309 
were performed with the High Energy Stereoscopic System (HESS)
from 2004 through 2007. Simultaneous optical/UV and X-ray observations 
were made with the XMM-Newton satellite on June 12/13 and June 14/15, 2005. NRT radio observations
were also contemporaneously performed in 2005.  ATOM optical monitoring observations
were also made in 2007.}
  {A strong VHE signal, $\sim$13$\sigma$ total, was detected by HESS 
after the four years HESS observations (116.8 hrs live time). 
The integral flux above 240 GeV is I($>$240 GeV) = 
$(3.06 \pm 0.26_{\rm stat} \pm 0.61_{\rm syst}) \times 10^{-12}$ cm$^{-2}$\,s$^{-1}$,
corresponding to $\sim$1.6\% of the flux observed from the Crab Nebula.
A time-averaged energy spectrum is measured 
from 200 GeV to 2 TeV and is characterized
by a power law (photon index of 
$\Gamma = 3.06 \pm 0.15_{\rm stat} \pm 0.10_{\rm syst}$).
Significant small-amplitude variations in the VHE flux from H\,2356$-$309
are seen on time scales of months and years, but not on shorter
time scales.  No evidence for any variations in the VHE spectral slope
are found within these data. The XMM-Newton X-ray measurements show a historically
low X-ray state, characterized by a hard, broken-power-law spectrum on both nights. }
  {The broad-band spectral energy distribution (SED) of the blazar can be adequately fit using a
simple one-zone synchrotron self-Compton (SSC) model.   
In the SSC scenario, higher VHE fluxes could be expected in the future since
the observed X-ray flux is at a historically low level.}


  \maketitle

\section{Introduction}

H\,2356$-$309 was suggested
by \cite{Costamante_AGN} as a strong candidate for the detection of VHE emission,
based on simple one-zone synchrotron self-Compton (SSC) modeling
of its spectral energy distribution (SED).
Motivated by these predictions, the HESS collaboration observed 
this blazar in 2004, resulting in the
discovery of VHE $\gamma$-ray emission (\cite{HESS_discovery}). 
Although several blazars are known to have variable 
VHE $\gamma$-ray fluxes (see, e.g., \cite{2155_flare}) 
and photon spectra (see, e.g., \cite{HESS_421}), no strong
evidence was found for such variability in the HESS data on
this source from June through December 2004. 

Hosted by an elliptical galaxy at a redshift of $z=0.165$
(\cite{redshift}), H\,2356$-$309 is among the more distant blazars 
detected at VHE energies. As a result, its observed spectrum is 
expected to be strongly affected (softened) by the absorption of
VHE photons on the extragalactic background light (EBL).  
The unexpectedly hard ($\Gamma \sim 3.1$) VHE spectrum 
measured by HESS in 2004 was used to confirm the 
strong constraints on the density of the EBL, near lower limits
based on the integrated light of resolved galaxies (\cite{HESS_discovery}),
given by the HESS spectrum from another VHE blazar (1ES\,1101$-$232).

At X-ray energies, H\,2356$-$309 was initially detected by
the UHURU satellite \cite{UHURU_det} and later by the HEAO-I satellite \cite{HEAO_det}.  
It is among the X-ray brightest of the 150 high-frequency peaked BL Lac objects (HBL)
in the Sedentary Survey \cite{Sedentary}, which contains
HBLs characterized by their exceptionally high X-ray-to-radio flux ratio.
BeppoSAX observations \cite{SAX_data} of H\,2356$-$309 in June 1998 
yielded an X-ray spectrum extending up to 50 keV, characterized by a broken power-law 
shape ($\Gamma_1 = 0.78^{+0.06}_{-0.07}$; $\Gamma_2 = 1.10 \pm 0.04$)
with a break at $1.8 \pm 0.4$ keV.  A softer power-law spectrum 
between 2 and 10 keV with an X-ray photon index of ($\Gamma_X$ = 2.43 $\pm$ 0.11)
was measured (\cite{HESS_MWLI}) during observations by 
the RXTE satellite on November 11, 2004.  The RXTE 
spectrum is softer above 2 keV and the flux
($9.7^{+0.3}_{-1.3} \times 10^{-12}$ erg cm$^{-2}$ s$^{-1}$)
is $\sim$3 times lower than that measured by BeppoSAX 
($2.5 \times 10^{-11}$ erg cm$^{-2}$ s$^{-1}$) in the same
energy band.

The 2004 RXTE data were taken simultaneously with HESS VHE,
ROTSE-IIIc optical, and Nancay radio-telescope observations.
The double-peaked SED derived from this campaign (\cite{HESS_MWLI}) 
was the first to include a contemporaneous VHE detection, and hence the
first to sample the higher-energy SED peak for this object since H\,2356$-$309 was not
detected by the EGRET satellite.  Modeling of these multi-wavelength observations
demonstrated that the SED of H\,2356$-$309 could be adequately fit 
by a simple one-zone stationary SSC scenario. 

Results from 76.9 hrs of new HESS observations of H\,2356$-$309
in 2005, 2006 and 2007 are reported here.  In addition, a re-analysis 
of the previously published 2004 HESS data (39.9 hrs), 
with an improved calibration of 
the absolute energy scale of the detector, is presented.
This improvement in the HESS calibration was not included in the previously 
mentioned modeling of the 2004 multi-wavelength observations,
therefore this modeling is repeated.  In addition,
the XMM-Newton satellite simultaneously observed H\,2356$-$309
on two nights ($\sim$5 hours each night) at X-ray energies, as well as at
optical through ultraviolet wavelengths, during the HESS observations in June 2005. 
The results from this multi-wavelength observation campaign, which
has much improved X-ray and optical measurements, are also discussed. 

\section{HESS Observations and Analysis Technique}

From 2004 through 2007, a total of 175.3 hours of HESS observations 
were taken
on H\,2356$-$309 in {\it wobble mode}, where 
the telescope tracking position is offset by $\pm$0.5$^{\circ}$ in either right
ascension or declination relative to the source location. 
After applying the standard HESS data-quality 
selection, the 
exposure is 116.8 hrs live time at 
a mean zenith angle of $Z_{\rm mean} = 19^{\circ}$.
The HESS data are calibrated following the procedures described in \cite{calib_paper}.
The analysis is performed using the standard HESS tools 
and $\gamma$-ray-like events are selected
using the {\it standard cuts} (\cite{std_analysis, HESS_crab}).
The average post-analysis energy threshold\footnote{The threshold is
corrected to account for the decreased 
optical efficiency of the HESS mirrors over the observation period.} is 240 GeV at $Z_{\rm mean}$.
The signal (on-source data) is taken from 
events with reconstructed direction falling within a circular region
of radius $\theta_{cut}$=0.11$^{\circ}$ centered
on H\,2356$-$309. The background (off-source data) is estimated using
the {\it Reflected-Region} method \cite{bgmodel_paper}, 
hence the on-source and off-source data are measured 
simultaneously from events in the same field of view.
Equation 17 in \cite{lima} is used
to calculate the significance of any excess.

The optical throughput of HESS decreases over time due
to degradation of the reflective surfaces of the HESS mirrors and Winston
cones, as well as to the accumulation of dust on the optical elements.
Compared to a newly-commissioned HESS telescope (mirror installation between
October 2001 and August 2003), this 
decrease is on average $\sim$26\% for the entire data sample,
and is 20\%, 27\%, 32\% and 35\% for
the H\,2356$-$309 data in 2004, 2005, 2006 and 2007, respectively.
This decrease causes images from gamma rays of equivalent energy to
appear less bright in later data, requiring a renormalization of the event energy.
This normalization
is made using the ratio of efficiencies determined from
simulated and observed muons (\cite{HESS_crab}), and 
eliminates long-term variations in the absolute energy scale
of the HESS analysis due to a changing optical throughput.

\section{HESS Results}

 \begin{table*}
  \begin{minipage}[t]{2.0\columnwidth}
     \caption{Results from four years of HESS observations of H\,2356$-$309.}
        \label{results}
       \centering
        \begin{tabular}{c c c c c c c c c c c c c}
           \hline\hline
           \noalign{\smallskip}
           Dark & MJD & MJD & Time & On & Off & $\alpha$ & Excess & Sig 
& I($>$240 GeV)\footnote{The 20\% systematic error on the observed integral flux above 240 GeV is not shown.} 
& Crab\footnote{The Crab Nebula flux percentage is
calculated relative to the HESS value above 240 GeV (\cite{HESS_crab}).} 
& $\chi^2$\,,\,NDF\footnote{The $\chi^2$, degrees of 
freedom (NDF), and corresponding $\chi^2$ probability P($\chi^2$) are given
       for fits of a constant to I($>$240 GeV) binned
       nightly within a dark period, or monthly within a year,
       or yearly within the total.} & P($\chi^2$)$^{c}$\\
           Period & First & Last & [hrs] & & & & &  [$\sigma$] & [10$^{-12}$\,cm$^{-2}$\,s$^{-1}$] & \% & & \\
           \noalign{\smallskip}
           \hline
           \noalign{\smallskip}
           06/2004 & 53172 & 53186 & 7.9  & 835 & 7778 & 0.0938 & 106 & 3.7 & $3.86 \pm 0.88$   & 2.1 & 8.9\,,\,11 & 0.63\\
           07/2004 & 53201 & 53209 & 3.0  & 280 & 2655 & 0.0938 & 31  & 1.8 & $1.19 \pm 1.18$   & 0.6 & 1.4\,,\,4 & 0.84\\
           09/2004 & 53256 & 53269 & 6.9  & 759 & 6015 & 0.0937 & 195 & 7.4 & $6.91 \pm 0.94$  & 3.7 & 4.0\,,\,6 & 0.67\\
           10/2004 & 53286 & 53296 & 9.9  & 950 & 8496 & 0.0914 & 174 & 5.7 & $4.33 \pm 0.80$   & 2.3 & 9.9\,,\,6 & 0.13\\
           11/2004 & 53316 & 53319 & 4.1  & 300 & 2700 & 0.0912 & 54  & 3.2 & $3.82 \pm 1.22$   & 2.0 & 7.0\,,\,3 & 0.07\\
           12/2004 & 53341 & 53353 & 8.2  & 559 & 5867 & 0.0926 & 16  & 0.6 & $1.23 \pm 1.46$   & 0.7 & 7.2\,,\,5 & 0.21\\
           06/2005 & 53530 & 53541 & 20.4 & 1200 & 12275 & 0.0921 & 69  & 2.0 & $1.78 \pm 1.04$   & 1.0 & 6.3\,,\,9 & 0.71\\
           07/2005 & 53556 & 53570 & 16.5 & 1455 & 13375 & 0.0922 & 221 & 5.9 & $3.08 \pm 0.60$   & 1.6 & 12.9\,,\,13 & 0.46\\
           08/2005 & 53583 & 53587 & 7.3  & 495 & 4920 & 0.0927 & 39  & 1.7 & $0.56 \pm 0.85$   & 0.3 & 5.7\,,\,4 & 0.22\\
           09/2005 & 53614 & 53615 & 2.6  & 152 & 1477 & 0.0946 & 12  & 1.0 & $2.70 \pm 1.67$   & 1.4 & 0.0\,,\,1 & 0.96\\
           07/2006 & 53937 & 53942 & 8.1  & 598 & 5926 & 0.0893 & 69  & 2.8 & $2.27 \pm 0.89$   & 1.2 & 2.5\,,\,5 & 0.77\\
           08/2006 & 53970 & 53978 & 2.2  & 188 & 1427 & 0.0894 & 60  & 4.8 & $7.39 \pm 1.84$  & 3.9 & 0.5\,,\,3 & 0.91\\
           09/2006 & 53999 & 54005 & 12.9 & 626 & 6044 & 0.0931 & 64  & 2.7 & $1.26 \pm 0.77$   & 0.7 & 7.2\,,\,6 & 0.30\\
	    07/2007 & 54295 & 54307 & 7.0  & 502 & 4670 & 0.0915 & 75  & 3.4 & $3.48 \pm 1.07$ & 1.9 & 8.9\,,\,10 & 0.54\\
          \noalign{\smallskip}
           \hline
           \noalign{\smallskip}
           2004 & 53172 & 53353 & 39.9 & 3683 & 33511 & 0.0927 & 576 & 9.6 & $4.06 \pm 0.42$ & 2.2 & 19.1\,,\,5 & 0.002\\
           2005 & 53530 & 53615 & 46.7 & 3302 & 32047 & 0.0924 & 341 & 5.9 & $2.22 \pm 0.44$ & 1.2 & 6.1\,,\,3 & 0.11\\
           2006 & 53937 & 54005 & 23.2 & 1412 & 13397 & 0.0910 & 193 & 5.1 & $2.36 \pm 0.56$ & 1.3 & 9.5\,,\,2 & 0.01\\
	    2007 & 54295 & 54307 & 7.0 & 502 & 4670 & 0.0915 & 75 & 3.4 & $3.48 \pm 1.07$ & 1.9 & $-$ & $-$ \\
           \noalign{\smallskip}
           \hline
           \noalign{\smallskip}
           Total & 53172 & 54307 & 116.8 & 8899 & 83625 & 0.0923 & 1185 & 12.6 & $3.06 \pm 0.26$ & 1.6 & 11.0\,,\,3 & 0.01\\
           \noalign{\smallskip}
           \hline
      \end{tabular}
   \end{minipage}
  \end{table*}

  \begin{figure}
  \centering
     \includegraphics[width=0.5\textwidth]{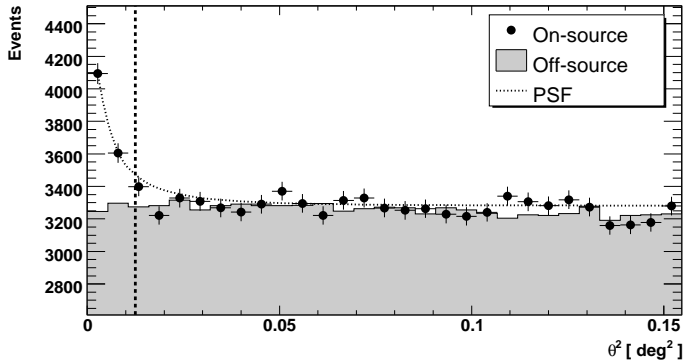}
     \caption{Distribution of $\theta^2$ for on-source 
       events (points) and
       normalized off-source events (shaded) from observations
       of H\,2356$-$309.  The dashed curve represents the
	expected distribution for a VHE $\gamma$-ray
	point source with photon index $\Gamma=3.06$ at a zenith angle of 20$^{\circ}$.  
	The vertical line denotes the on-source integration region relative
	to the nominal source position (\cite{redshift}).}
        \label{thtsq_plot}
  \end{figure}

A total of 1185 excess events (12.6 $\sigma$)
is measured from the direction of H\,2356$-$309 in the complete data set.
Figure~\ref{thtsq_plot} shows the on-source and normalized off-source
distributions of the square of the angular difference between
the source position (\cite{redshift}) and
the reconstructed shower position ($\theta^{2}$) for all observations. 
The observed signal can be seen at small values of
$\theta^{2}$ and the distribution of the excess is similar to
that expected from a simulated $\gamma$-ray point source.

H\,2356$-$309 is clearly detected with a statistical significance
of more than 5 standard deviations ($\sigma$), in each of the first three
years (2004-06) of HESS observations, and a marginal excess (3.4 $\sigma$)
is observed during the brief observations in 2007. 
Detailed results of the HESS observations of H\,2356$-$309 for individual
observation periods are summarized in Table~\ref{results}.  The table 
contains the dead-time-corrected observation time, 
the number of on and off-source events, the on/off normalization ($\alpha$),
the observed excess and its statistical significance for various intervals (e.g. dark periods) 
in which H\,2356$-$309 was observed.

A two-dimensional Gaussian fit to the sky map of the observed excess demonstrates that
the excess is point-like with an upper limit (99\% confidence level) on
the extension of 1.3'.   The centroid of the fit, HESS\,J2359$-$306, is located at
$\alpha_{\rm J2000}=23^{\mathrm h}59^{\mathrm m}7.8^{\mathrm s} \pm 1.6^{\mathrm s}_{\rm stat} \pm 1.3^{\mathrm s}_{\rm syst}$
and $\delta_{\rm J2000}=-30^{\circ}37'18'' \pm 20''_{\rm stat} \pm 20''_{\rm syst}$. HESS\,J2359$-$306 is
spatially consistent with the position (\cite{redshift}) of the blazar 
($\alpha_{\rm J2000}=23^{\mathrm h}59^{\mathrm m}7.8^{\mathrm s}$, 
$\delta_{\rm J2000}=-30^{\circ}37'38''$) as expected. The distance between the two positions is 
$20'' \pm 31''$.

\subsection{VHE Spectrum of H\,2356$-$309}

  \begin{figure}
  \centering
     \includegraphics[width=0.5\textwidth]{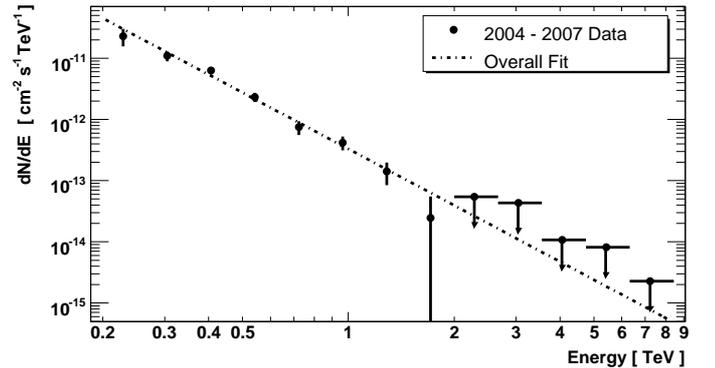}
     \caption{Time-averaged (2004-2007) VHE photon spectrum of H\,2356$-$309. 
	The dashed line represents the best $\chi^2$ fit of a power law to
        the data up to 2 TeV, and a linear extrapolation at higher energies.
	The upper limits are at the 99\% confidence
	level \cite{UL_tech}. Only the statistical errors are shown.}
        \label{spectrum_plot}
  \end{figure}

  \begin{figure}
  \centering
     \includegraphics[width=0.5\textwidth]{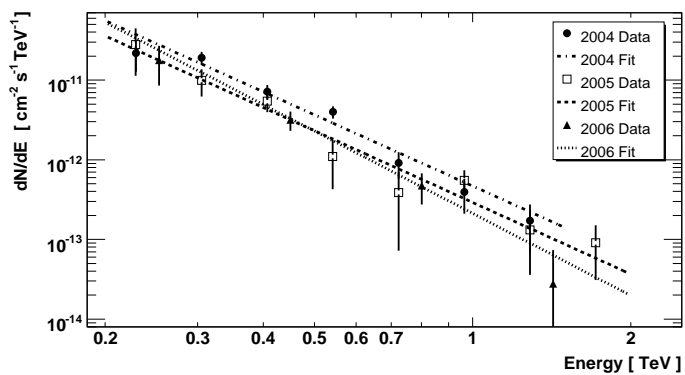}
     \caption{Annual VHE photon spectra observed by HESS 
	from H\,2356$-$309. Each line represents the best 
	$\chi^2$ fit of a power law to the observed data. Only the statistical errors are shown.}
        \label{annual_spectra}
  \end{figure}

  \begin{table*}
    \begin{minipage}[h]{2.0\columnwidth}
     \caption{Results of power-law fits to the VHE spectra of H\,2356$-$309.}
        \label{annual_results}
       \centering
        \begin{tabular}{c c c c c c c c}
           \hline\hline
           \noalign{\smallskip}
            Epoch & E$_{\rm min}$ & E$_{\max}$ & $\Gamma$ & $I_{\circ}$\footnote{The differential flux normalization is calculated at 1 TeV.} & $\chi^2$ & NDF & P($\chi^2$)\\
		& [TeV] & [TeV] & & [$10^{-13}$ cm$^{-2}$\,s$^{-1}$\,TeV$^{-1}$] & & & \\
           \noalign{\smallskip}
           \hline
           \noalign{\smallskip}
	    2004 (AH06)\footnote{ The AH06 entry is the
	previously published HESS result (\cite{HESS_discovery})
	for the 2004 data and is not corrected for
	long-term changes in the optical efficiency of the system.}
& 0.165 & 1.041 & $3.06 \pm 0.21_{\rm stat} \pm 0.10_{\rm syst}$ & $3.08 \pm 0.75_{\rm stat} \pm 0.62_{\rm syst}$ & 3.9 & 6 & 0.69 \\
           \noalign{\smallskip}
           \hline
           \noalign{\smallskip}
	    2004\footnote{This 2004 entry is derived from the exact same data as presented in AH06,
	but with a correction (see text) applied to account for
	the optical efficiency changes within the data. The 2005, 2006 and
	total entries also have this correction applied.}
& 0.200 & 1.500 & $2.97 \pm 0.19_{\rm stat} \pm 0.10_{\rm syst}$ & $4.69 \pm 0.86_{\rm stat} \pm 0.94_{\rm syst}$ & 7.8 & 5 & 0.17\\
	    2005$^{\hspace{0.1cm}}$ & 0.200 & 2.000 & $2.99 \pm 0.39_{\rm stat} \pm 0.10_{\rm syst}$ & $2.92 \pm 0.89_{\rm stat} \pm 0.58_{\rm syst}$ & 5.0 & 6 & 0.55\\
	    2006\footnote{Due to low statistics, the 2006 spectrum
       is determined using 4 bins/decade in energy instead of the
	8 bins/decade used for the other photon spectra.} & 0.200 & 2.000 & $3.43 \pm 0.41_{\rm stat} \pm 0.10_{\rm syst}$ & $2.13 \pm 0.79_{\rm stat} \pm 0.43_{\rm syst}$ & 0.7 & 2 & 0.70\\
           \noalign{\smallskip}
           \hline
           \noalign{\smallskip}
	    Total & 0.200 & 2.000 & $3.06 \pm 0.15_{\rm stat} \pm 0.10_{\rm syst}$ & $3.29 \pm 0.45_{\rm stat} \pm 0.66_{\rm syst}$ & 5.7 & 6 & 0.47\\
         \noalign{\smallskip}
           \hline
      \end{tabular}
    \end{minipage}
  \end{table*}

Figure~\ref{spectrum_plot} shows the four-year (2004--2007) time-averaged photon spectrum
of H\,2356$-$309.  The best $\chi^2$ fit of a power law
(dN/dE = $I_{\circ} (E/{\rm 1\,TeV})^{-\Gamma}$) to 
these data yields a photon index 
$\Gamma=3.06 \pm 0.15_{\rm stat} \pm 0.10_{\rm syst}$, 
and a $\chi^2$ of 5.7 for 6 degrees of freedom.
The last spectral point at $\sim$1.7 TeV is not significant (0.8 $\sigma$),
however removing it from the fit does not appreciably change the result.
Fitting the observed photon spectrum with a more complex function,
such as a power-law with a cutoff or break, does not yield a $\chi^2$
that is significantly improved (as determined by an F-test).

The high photon statistics observed in 2004, 2005 and 2006 enable
the determination of a spectrum for each of those years.  Unfortunately
a spectrum could not be generated for the 2007 data due to low statistics.
The annual spectra for the first three years of observations are shown
in Figure~\ref{annual_spectra} together with the best $\chi^2$ fit
of a power law to the data.  Table~\ref{annual_results}
gives the fit results for each of the annual spectra, as well as for the total
spectrum. The epoch, lower and upper energy bounds,
photon index ($\Gamma$), differential flux 
normalization at 1 TeV ($I_{\circ}$), $\chi^2$, 
degrees of freedom (NDF), and $\chi^2$ probability (P($\chi^2$))
for each fit are shown.  Although the differential
flux normalization $I_{\circ}$ is $\sim$1.5 times larger in 2004 than in
2005 and 2006, there are no significant changes in the spectral slope
during the HESS observations.

\subsection{VHE Flux from H\,2356$-$309}

  \begin{figure}
  \centering
     \includegraphics[width=0.5\textwidth]{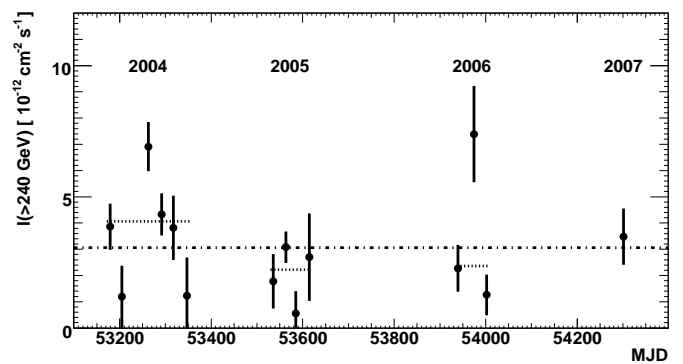}
     \caption{Integral flux, I($>$240 GeV), measured by HESS
from H\,2356$-$309 during each dark period of observations.
For each point the time-averaged $\Gamma=3.06$ 
is assumed and only the statistical errors are shown.
The horizontal line represents the average flux for all 
the HESS observations.  The three horizontal line segments are
the average 2004, 2005 and 2006 fluxes.}
        \label{monthly_plot}
  \end{figure}

\begin{figure*}
  \centering
     \includegraphics[width=0.6\textwidth]{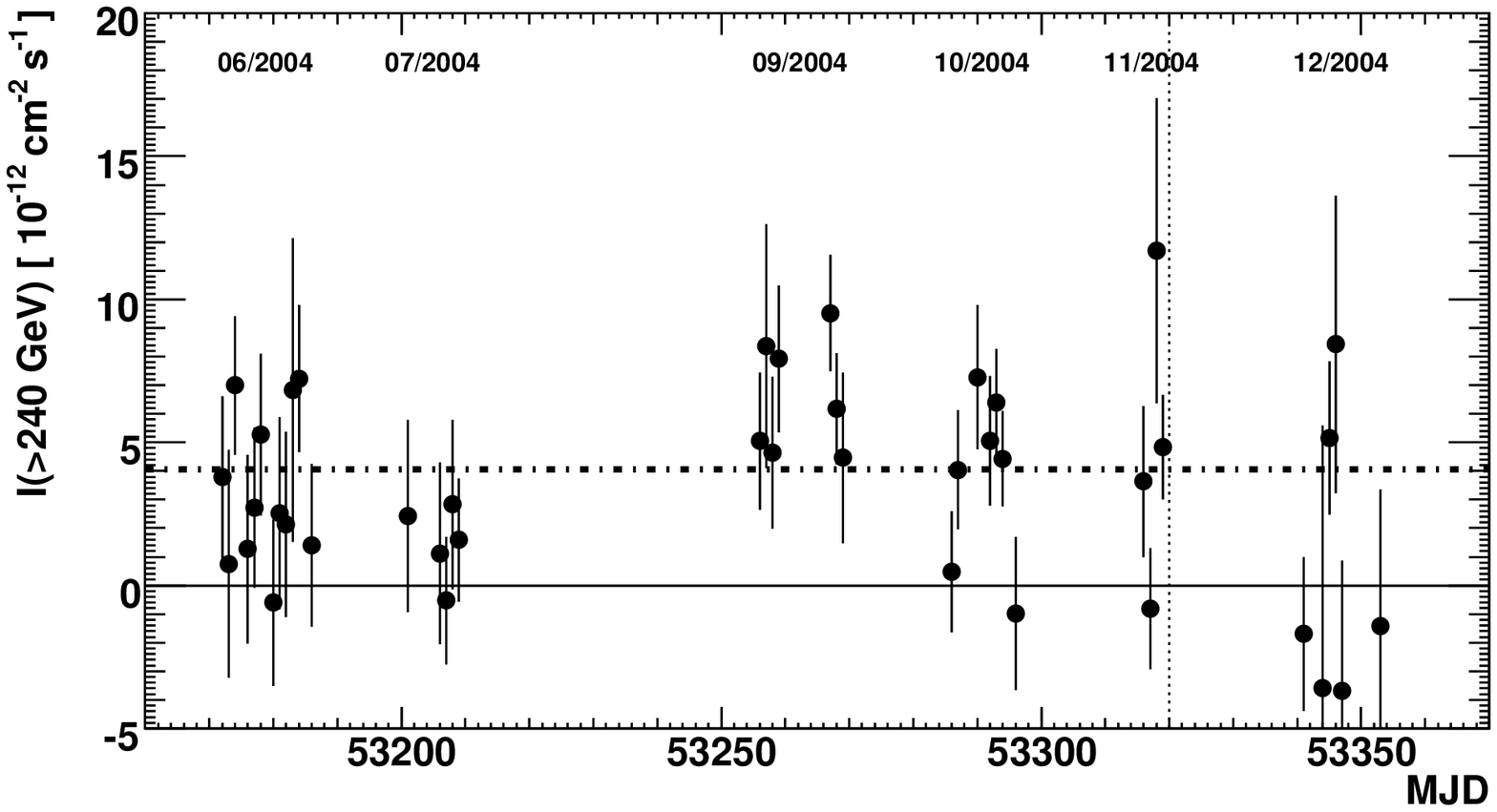}\\
     \includegraphics[width=0.6\textwidth]{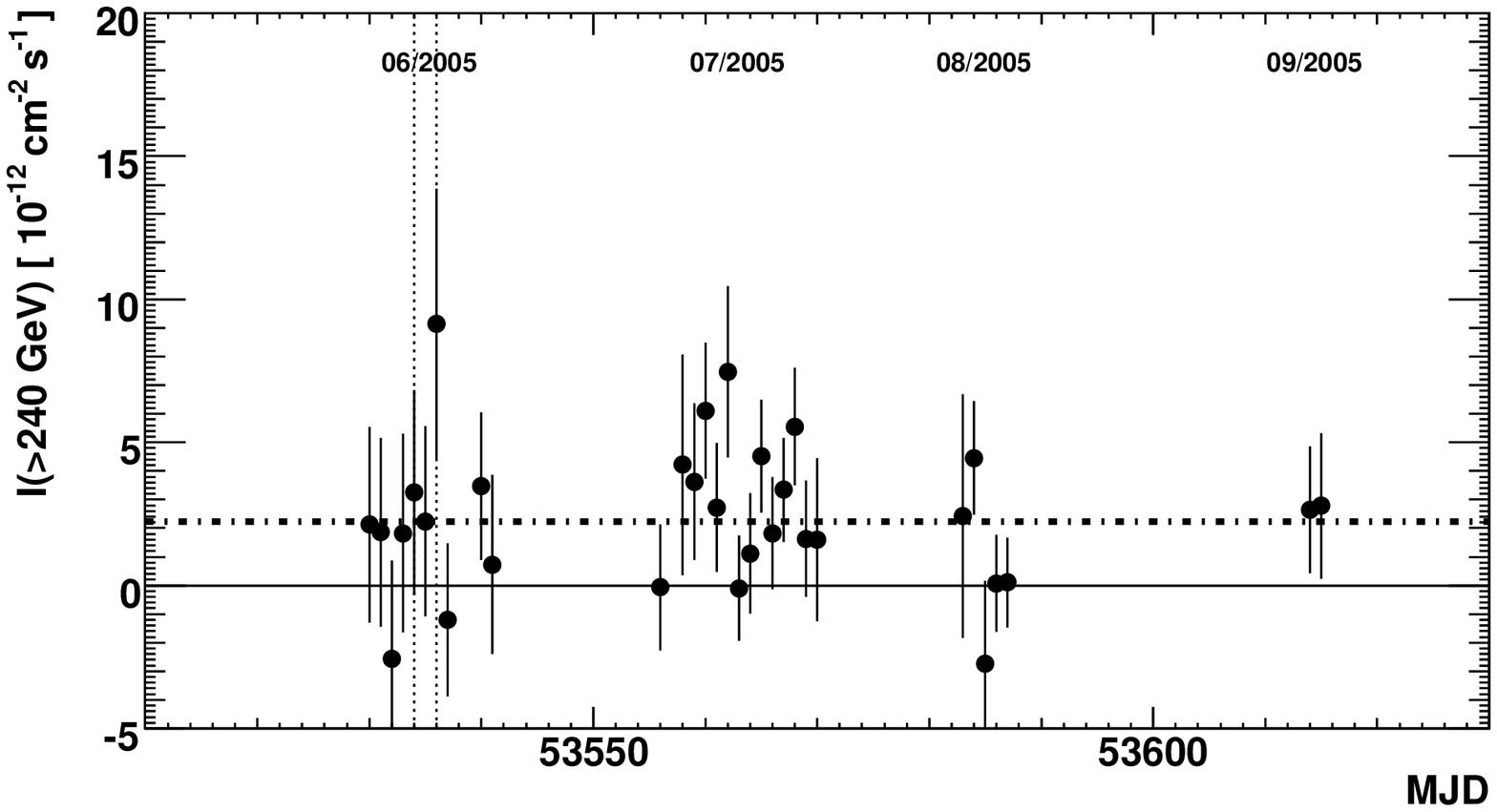}\\
     \includegraphics[width=0.6\textwidth]{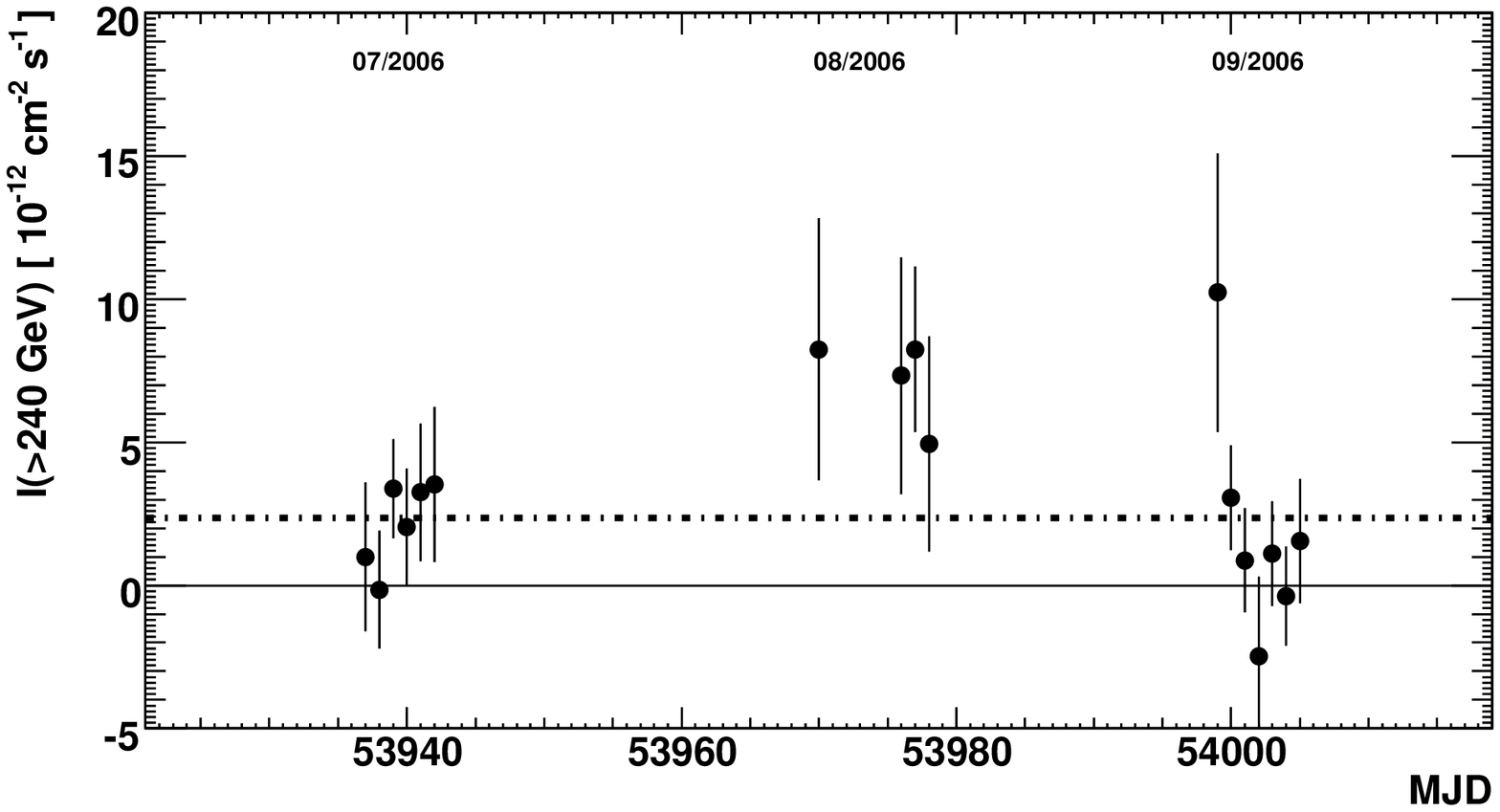}\\
     \includegraphics[width=0.6\textwidth]{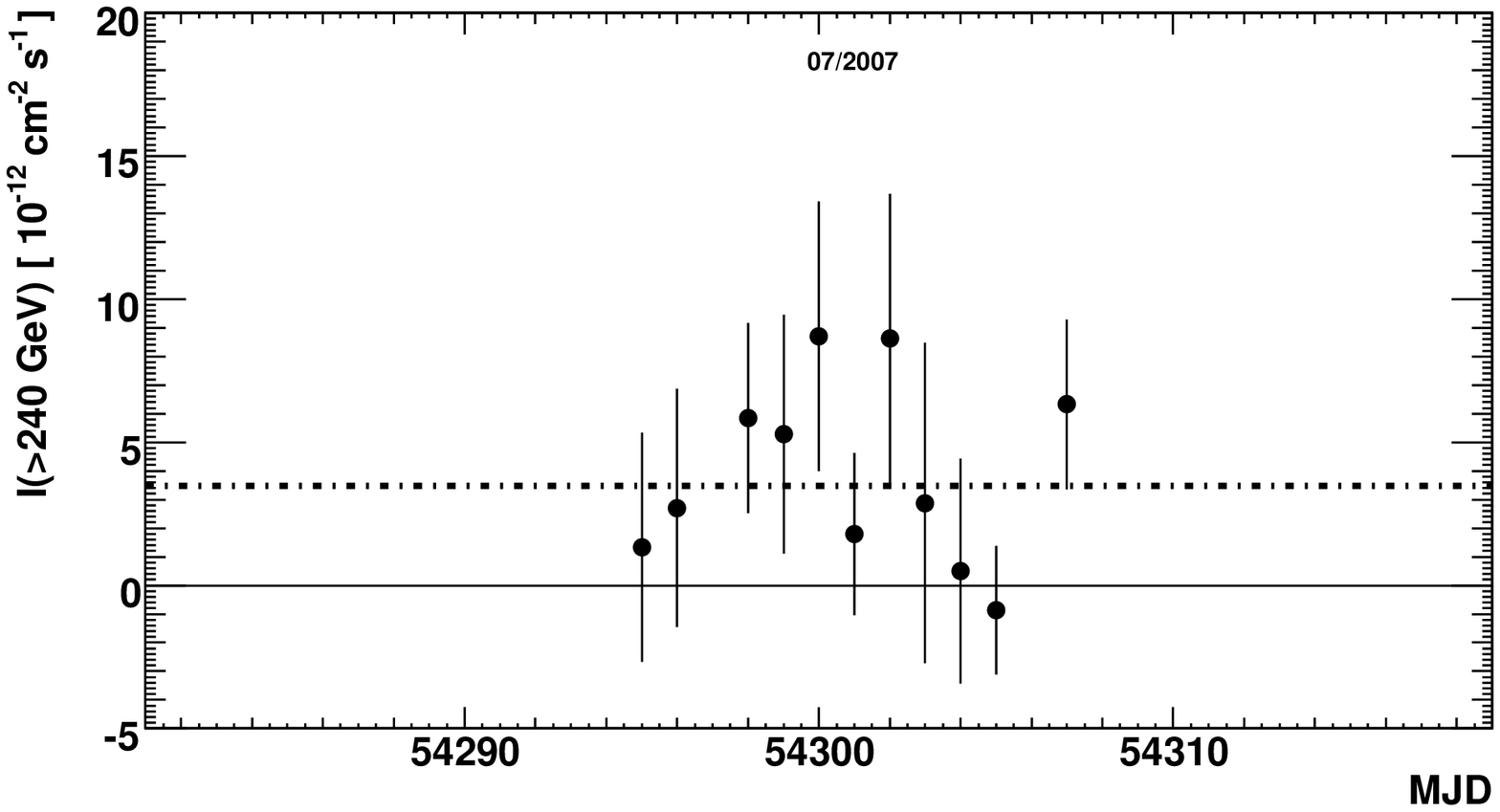}\\
     \caption{Integral flux, I($>$240 GeV), measured by HESS
from H\,2356$-$309 during each night of observations in 2004 (top), 2005 (top-middle), 
2006 (bottom-middle) and 2007 (bottom).
For each point the time-averaged $\Gamma=3.06$ is assumed, and
only the statistical errors are shown.  The horizontal
line in each figure represents the average flux measured by
HESS in the respective year. The night of the RXTE observations (MJD 53320)
is denoted by the dotted vertical line in the 2004 light curve.  The nights (MJD 53534 \& 53536) of
the XMM-Newton observations are marked by dotted vertical lines in the 2005 figure.}
        \label{nightly_plots}
  \end{figure*}

The observed integral flux\footnote{All HESS fluxes (i.e. annual, dark period and nightly 
values) in this article are
calculated using the time-averaged $\Gamma=3.06$.} above 240 GeV for the entire data set is
I($>$240 GeV) = $(3.06 \pm 0.26_{\rm stat} \pm 0.61_{\rm syst}) \times 10^{-12}$ 
cm$^{-2}$\,s$^{-1}$, corresponding to $\sim$1.6\% of I($>$240 GeV)
determined by HESS from the Crab Nebula (\cite{HESS_crab}).
Figures \ref{monthly_plot} and \ref{nightly_plots} 
show the flux measured for each dark period and night, respectively.
The integral flux I($>$240 GeV)
for each year of observations, as well as for each dark period,
is shown in Table~\ref{results}. Also given are the $\chi^2$ and corresponding probability 
for a fit of a constant to the data when binned by nights within each dark period, 
by dark periods within a year, and by year within the total observations.  There are
clear indications that the VHE flux from H\,2356$-$309 varies weakly on
time scales of months and years.  There is no evidence
for flux variability on any shorter time scale within the HESS
data, however the average flux is too low to significantly detect similar weak
variations during such brief exposures.

\subsection{Optical Efficiency Correction}

\begin{figure}
  \centering
     \includegraphics[width=0.5\textwidth]{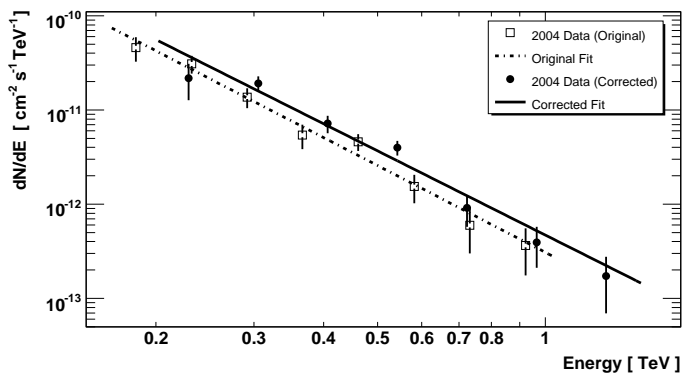}
     \caption{Renormalized VHE spectrum measured by HESS from H\,2356$-$309 in
	2004 compared to the previously published version 
	(\cite{HESS_discovery}). Only the statistical errors are shown.}
        \label{muoncorr_plot}
  \end{figure}

The previously published (\cite{HESS_discovery})
spectrum of H\,2356$-$309 in 2004 was
not corrected for optical sensitivity changes.
Figure ~\ref{muoncorr_plot} illustrates
the effect of correcting the energy of individual events for the relative 
optical efficiency of the system.  The results
of the best fit to each set of points is shown in  Table~\ref{annual_results}.
As can be seen, the corrected  spectrum has a significantly larger
flux normalization ($I_{\circ}$), but the photon index ($\Gamma$)
is unchanged.  The corrected 2004 integral flux 
is $\sim$50\% higher than previously published.  The increase in
integral flux due to the optical efficiency correction is relatively
large for H\,2356$-$309 because of the softness of the observed VHE specrum.
The correction is smaller for harder spectrum sources 
(e.g., typical Galactic VHE sources).

\section{Multi-wavelength Observations}
The emission from blazars is known to span over twenty orders of
magnitude in energy and is variable at all energies.  
Accurate modeling of the underlying
processes in these objects thus requires
simultaneous observations at many wavelengths.
Two such simultaneous multi-wavelength observation
campaigns were organized by the HESS collaboration
for H\,2356$-$309.  The first of these included
the RXTE X-ray satellite, the ROTSE-IIIc optical
telescope and the Nancay Radio Telescope (NRT), and is
henceforth referred to as the RXTE campaign.
The second campaign involved X-ray and optical/UV observations
with XMM-Newton satellite and further NRT data, and is henceforth
referred to as the XMM-Newton campaign.  In addition, results 
of ATOM optical monitoring of H\,2356$-$309 beginning in
November 2006 are presented.

\subsection{RXTE Campaign}
Results from the RXTE campaign were first reported in
Aharonian et al. (2006b).  As only the normalization of
the HESS observations have changed, the previously reported lower-frequency results
are summarized here.  The RXTE/PCA observed H\,2356$-$309
for a total of 2.14 ks on November 11, 2004 (MJD 53320) as a target-of-opportunity
request.  The constant
X-ray flux was $9.7^{+0.3}_{-1.3} \times 10^{-12}$ erg cm$^{-2}$ s$^{-1}$
in the 2$-$10 keV band.  The X-ray spectrum was generated by combining data
from three separate PCUs, yielding a single-PCU equivalent exposure of 5.42 ks,
and is compatible with a power-law function with photon index $\Gamma_X=2.43 \pm 0.11$.
The average NRT 11-cm flux between June 11 and October 10, 2004 was
$40 \pm 8$ mJy.  During the observations by HESS in 2004, 
ROTSE-IIIc measured apparent R-band magnitudes $m(R)$ between 16.1 and 16.9, with
an estimated contribution from the resolved elliptical host galaxy of $m(R)=17$.
Due to poor weather, HESS was unable to observe the blazar on MJD 53320.
However, there is no evidence for variability in the November 2004
VHE flux and this flux is consistent with the average 2004 VHE flux.
As there is no evidence for changes in the spectral slope of H\,2356$-$309 on any time
scale, the time-averaged VHE spectrum from 2004 is assumed to represent
the VHE state of the blazar during the RXTE campaign for the SED modeling
in Section~\ref{Discuss}.  The same assumption was made in 
Aharonian et al. (2006b).

\subsection{XMM-Newton Campaign}
The XMM-Newton satellite was pointed at H\,2356$-$309
on two nights in 2005. The first XMM-Newton pointing
began at 23:59 UTC on June 12, 2005 and ended at 5:17 UTC
on June 13, 2005 (MJD 53334).  The second pointing started at
23:57 UTC on June 14, 2005 and finished at 5:19 UTC
on June 15, 2005 (MJD 53536).  Simultaneous HESS observations of the blazar 
were also made during both of the XMM-Newton pointings.
At radio wavelengths, the 11-cm flux from H\,2356$-$309 was 
contemporaneously monitored by the NRT.

\subsubsection{HESS Results}

The total HESS exposure during the XMM-Newton campaign
is 4.68 hrs live time after selection criteria.  The data
divided approximately evenly over the two nights with exposures of 2.44 hrs and 2.24
hrs on MJD 53534 and 53536, respectively.
HESS did not detect a significant excess from H\,2356$-$309
on either night of the XMM-Newton observations.  During these two nights, 
a total of 302 on-source events
and 2907 off-source events were measured with an 
on-off normalization of 0.0930,
corresponding to an excess of 32 events (1.8$\sigma$).
Due to the lack of a significant detection 
it is not possible to produce a photon spectrum from the data.
Assuming the observed excess is from $\gamma$-rays emitted by
H\,2356$-$309, the average flux during the XMM-Newton observations,
I($>$240 GeV) = $(5.7 \pm 2.9_{\rm stat} \pm 1.1_{\rm syst}) \times 10^{-12}$
cm$^{-2}$\,s$^{-1}$, is consistent with time-averaged value from 2005.
The flux for each of the nights (MJD 53534 and
53536) in the XMM-Newton epoch can be seen in Figure~\ref{nightly_plots}.
There are no significant variations of the nightly flux, or of the
run-wise ($\sim$28 min) flux in either of the individual nights, during the XMM-Newton observations.
Since there are no indications of any VHE flux variations in 2005,
the time-averaged VHE spectrum from 2005 is used for the
SED modeling of the each night of the XMM-Newton campaign in Section~\ref{Discuss}.

\subsubsection{XMM-Newton X-ray Results}

   \begin{table*}[th]
     \begin{minipage}[t]{2.0\columnwidth}
       \caption{ Spectral information from X-ray observations of H\,2356$-$309. }
     \centering
      \begin{tabular}{lcccccccc}
       \hline
       \hline
            \noalign{\smallskip}
Instrument & MJD  &  Model & $\Gamma_1$ & $E_{\rm break}$    & $ 
\Gamma_2$ &$F_{0.1-2.0~\rm keV}$  & $F_{2-10~\rm keV}$ & $\chi^2_r$\,,\,NDF\footnote{The reduced $\chi^2_r$ and degrees of freedom (NDF)
correspond to fits of either a power-law (PL) function or broken power-law (BPL) function to the XMM-Newton X-ray data.}\\
            \noalign{\smallskip}
          &       &         &    & [keV]  &   &  \multicolumn{2}{c}{[$10^{-12}$ ergs cm$^{-2}$ s$^{-1}$]\footnote{The integrated fluxes have the effects of Galactic absorption removed.}} &  \\
            \noalign{\smallskip}
	     \hline
            \noalign{\smallskip}
BeppoSAX & 51350$-$51 & BPL & $0.78^{+0.06}_{-0.07}$ & $1.8 \pm 0.4$ &  
$1.10 \pm 0.04$ & $-$ & 25 & $-$\\
RXTE & 53320 & PL & $2.43 \pm 0.25$ & $-$ & $-$ & $-$ & $9.70^{+0.3}_ 
{-1.3}$ & $-$\\
            \noalign{\smallskip}
\hline
            \noalign{\smallskip}
XMM-Newton\footnote{All errors from the XMM-Newton fits are quoted at the 90\% confidence level.} & 53534 & PL & $2.23 \pm 0.01$ & $-$    &  $-$   & 25.9 & 8.05 &  1.41\,,\,541\\
    &      & BPL & $2.08 \pm 0.03$ & $1.00 \pm 0.08$ &  $2.32 \pm 0.02$ &  23.2 & 7.46 & 0.95\,,\,539 \\
            \noalign{\smallskip}
XMM-Newton$^{c}$ & 53536 & PL    & $2.16 \pm 0.01$ & $-$  &  $-$  & 25.9 & 9.54 &  1.52\,,\,541\\
    &       & BPL & $1.89 \pm 0.10$ & $0.66 \pm 0.14$ &  $2.20 \pm 0.02$ &  22.3 & 9.15 & 1.21\,,\,539\\
            \noalign{\smallskip}
       \hline
       \hline
     \end{tabular}
    \label{Xray_results}
  \end{minipage}
  \end{table*}

During the two pointings,
the XMM-Newton EPIC instruments were set in timing (PN and MOS1
cameras) and large window (MOS2 camera) modes.
On MJD 53534 the total live time of the XMM-Newton
exposures are $T_{\rm PN} = 16.5$ ksec for the EPIC PN instrument
and  $T_{\rm MOS2} = 17.6$ ksec for the EPIC MOS2 detector.
The live times are similar, $T_{\rm PN} = 16.6$ ksec
and  $T_{\rm MOS2} = 17.7$ ksec, on MJD 53536.
The XMM-Newton Science Analysis System (SAS version 7.1.0)
is used to process the data with the July 2008 calibration files.
The spectral and timing analysis are performed with {\tt XSPEC v11.3.2ag}
and {\tt FTOOLS V6.3.2}.
The MOS2 data are taken in Large Window mode, while 
the MOS1 in Timing mode.  Data from the MOS1 detector are not included,
as they are in qualitative agreement with the other instruments
but exhibit much stronger noise levels.  
The 0.1$-$1 keV and 4$-$10 keV count rates of the PN, as well as the
corresponding hardness ratio, are all constant in time on
each of the two nights. Therefore only the night-averaged
energy spectra are presented.

The effects of pile-up in the EPIC instruments were explored using
{\tt epatplot}.  No significant pile-up is present.
The spectrum is obtained by selecting only single pixel ({\tt PATTERN=0}) events
for the MOS2 data, and single plus double pixel 
events ({\tt PATTERN$\le$4}) for the PN data.

For the MOS2 analysis, the signal is extracted from a circle of radius
$45''$ centered on the source centroid.  The background is taken from an
annulus around the source with an inner radius of $120''$ and an outer radius
set by the border of the CCD window.   For the PN analysis,
signal photons are selected from rows $28 \le {\rm RAWX} \le 48$ and the
background is estimated from $2 \le {\rm RAWX} \le 18$.  The energy
range of the PN is restricted to 0.5 to 10 keV, while MOS2 events  
between 0.15 and 10 keV are accepted.  
For the spectral determination, different re-binning schemes
are used, always requiring at least 50 counts per new bin. 
Ancillary and response files are produced with {\tt rmfgen} and {\tt  
arfgen}.

The spectra of the PN and MOS2 detectors are combined in the following
fitting, with a free constant (that remained within a few percent of unity)
to allow for the different normalization between the two instruments.
The spectra are fit with source models including
Galactic absorption along the line of sight to
H\,2356$-$309. The absorption model {\tt TBabs} by \cite{wilms} is used,
with cross-sections 
by \cite{verner}.  The absorbing column density is kept fixed at
the Galactic value of $N_{\rm H} = 1.44 \times 10^{20}$ cm$^{-2}$ (\cite{gal_abs}).

A single power-law model does not provide an acceptable
fit to the X-ray data of either night (see Table \ref{Xray_results}).
Different models of the Galactic absorption ({\tt wabs}, {\tt phabs})  were also
tested, but none of them provide an acceptable fit for a single power-law 
source model. A broken power-law provides an improved fit on both  
nights (F-test $>99.999$\%), and also with respect to a fit with
free $N_{\rm H}$ (reduced $\chi^2_r=0.95$ for 539 NDF  vs. $\chi^2_r=1.12$ for 540 NDF).

\begin{figure*}
   \centering

  $\begin{array}{c@{\hspace{0.5cm}}c}

  \multicolumn{1}{l}{\mbox{\bf }} &
        \multicolumn{1}{l}{\mbox{\bf }} \\ [0cm]

  \includegraphics[width=0.33\textwidth, angle=270]{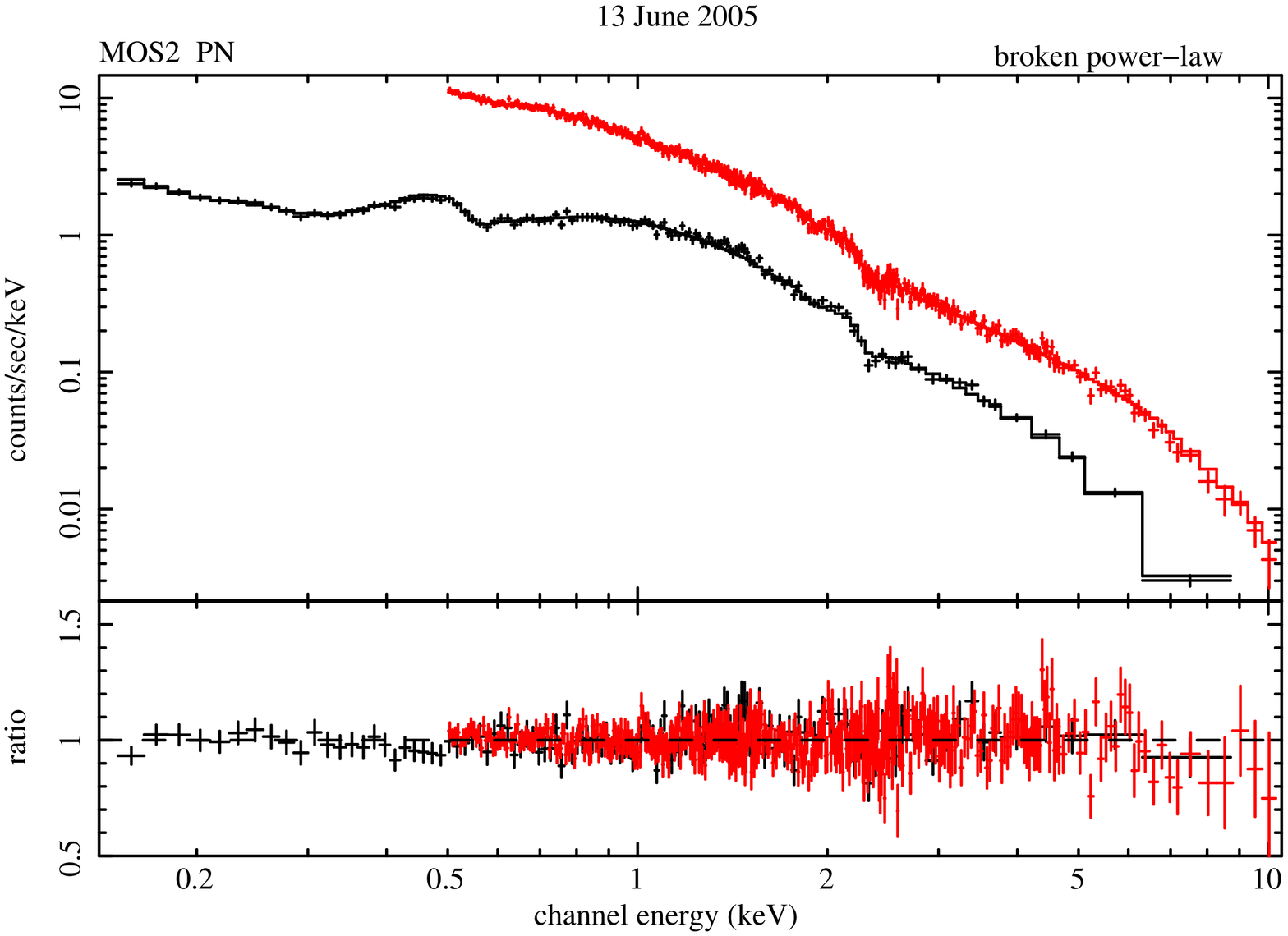} &
  \includegraphics[width=0.33\textwidth, angle=270]{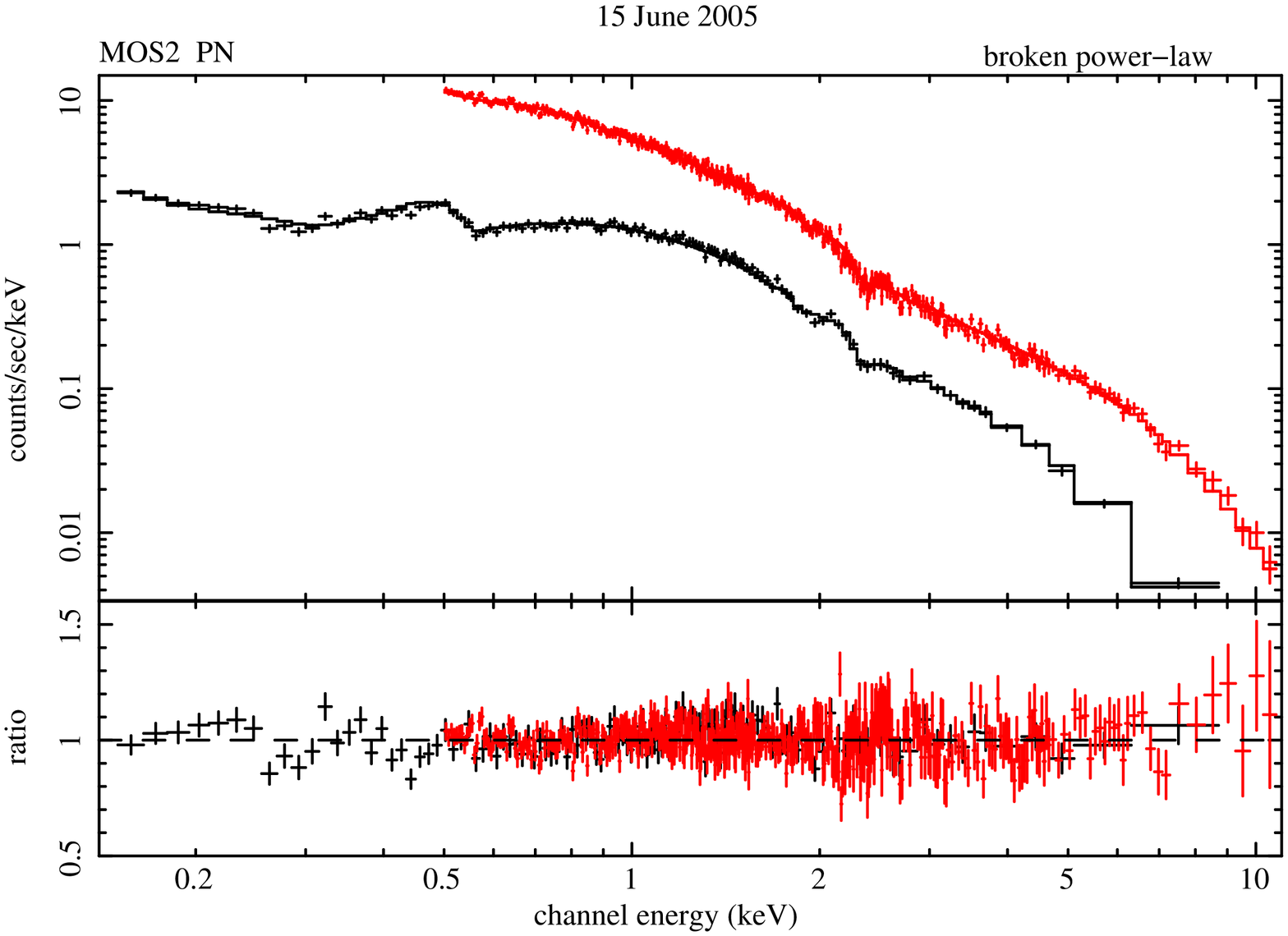} \\ [0.0cm]

  \end{array}$
      \caption{X-ray spectra measured during the XMM-Newton observations in 2005.  The lefthand and righthand
figures correspond to the measurements on MJD 53534 and 53536, respectively. The red points (0.5 to 10 keV) and
black points (0.15 to 10 keV) are data from the PN and MOS2 detectors, repectively.  The corresponding
fits to the data are also shown as lines, and the fit residuals are shown in the lower panels. }
         \label{Xray_spec}
   \end{figure*}

The results of the broken power-law fits for
both nights are shown in Table~\ref{Xray_results}, along with a  
summary of archival X-ray spectrum measurements.
The measured X-ray spectra for the XMM-Newton pointings on MJD 53534 and 53536
are shown in Figure~\ref{Xray_spec}.

\subsubsection{XMM-Newton Optical Monitor Results}

During the two pointings the XMM-Newton optical monitor (OM) took four exposures 
with four different filters (V, B, U, UVW1). For each
exposure, the window on the target was set in fast mode.
The XMM-Newton OM data are processed with {\tt xmmsas 7.1.0} using
data from the imaging mode for the photometry, and 
from the fast timing window for the light curves.
No variability is found in any of the OM exposures for each of the different filters.
The point-source analysis of the OM photometry program {\tt omsource} 
is used to extract count rates, which are
converted into fluxes following the standard procedures.
For the V, B, and U filters, the on-source data
are taken from an aperture of $R = 6''$ and the background
is estimated from an annulus with $10'' < R < 15''$.
An on-source aperture of  $R = 17''.5$ and background annulus
of $20'' < R < 25''$ are used for the UVW1 filter.
Consistent results were also obtained using different 
background regions. 

The optical spectrum for each of the two nights is generated by
combining the results from all four filters.  Each of the V and B band
exposures of H\,2356$-$309 contains a significant contribution
from the host galaxy that must be subtracted.  The host galaxy is
resolved at optical (\cite{redshift,optical_res}) and near-infrared
(\cite{optical_res2}) wavelengths, and is a normal elliptical galaxy
with an effective radius of about $1''.7$ in the V and R bands,
somewhat lower in the IR.  Using a standard de Vaucouleurs radial
profile, more than 60\% of the host-galaxy flux is thus estimated to be
contained in the V and B band signal apertures. The host-galaxy contribution to the
background apertures is negligible. These fractions of the host galaxy
flux are subtracted from the V and B band fluxes using the R band
magnitude ($m_{\rm R}$ = 17.21) from \cite{Urry_mag},
and the spectral template of elliptical galaxies ($z=2$) from \cite{spectral_template}.  
The host-galaxy-subtracted fluxes are corrected
for Galactic extinction using $A_B = 0.058$ mag and the interstellar
reddening curve by \cite{reddening}, updated by \cite{reddening_update}.
The exposure ($T$) and observed flux ($F$) in each optical band are
shown in Table~\ref{optical_results} for each of the two nights. The
contribution of the host galaxy to the other filters is negligible.

   \begin{table}
     \begin{minipage}[]{1.0\columnwidth}
      \caption{XMM -Newton Optical Monitor results.}
         \label{optical_results}
        \centering
         \begin{tabular}{c c c c c}
            \hline\hline
            \noalign{\smallskip}
 Band & T$_{1}$\footnote{The OM exposures (T) and fluxes (F) are reported
for each filter on MJD 53534 (T$_1$ \& F$_1$) and 53536 (T$_2$ \& F$_2$).} & F$_{1}$ & T$_{2}$$^a$ & F$_{2}$\\
 & [ksec] & [mJy]& [ksec] & [mJy]\\
            \noalign{\smallskip}
            \hline
            \noalign{\smallskip}
               V & 4.3 & 0.33 & 3.8 & 0.32\\
               B & 3.9 & 0.22 & 4.3 & 0.21\\
               U & 3.9 & 0.24 & 4.3 & 0.23\\
               UVW1 & 4.4 & 0.24 & 4.4 & 0.23\\
            \noalign{\smallskip}
            \hline
       \end{tabular}
     \end{minipage}
   \end{table}

\subsubsection{NRT Results}
The NRT (\cite{Nancay1}), a meridian transit telescope
with a main spherical mirror of 300 m $\times$ 35 m,
measured the 11-cm flux from H\,2356$-$309 
on 29 different days between 
June through December 2005 as part of an
on-going monitoring program.  The nearest observations
to the XMM-Newton pointings were performed on the three nights
of June 11$-$13, 2005. The observed 11-cm flux was quite low and 
not strongly variable during 2005, 
with an average of 7.5 $\pm$ 2.0 mJy.

\subsection{ATOM Results}
Optical observations were taken using the ATOM telescope \cite{ATOM} at the HESS
site from November 2006 on. Absolute flux values are calculated
using differential photometry against several stars listed in the
NOMAD catalog \cite{NOMAD}. A 4\,arcsec radius aperture
is used for all filter bands. A total of $\sim$500 measurements in
4 filter bands were taken between November 2006 and the end of 2007.  The
measured apparent magnitudes vary in the R band around ($16.4 \pm 0.2$) mag. 
For the other filter bands, the values are ($17.1\pm0.2$) mag in B, 
($17.0\pm0.2$) mag in V and ($16.1 \pm 0.2$) mag in I band.

\section{Discussion \label{Discuss}}

  \begin{figure*}
  \centering
     \includegraphics[width=0.7\textwidth, angle=270]{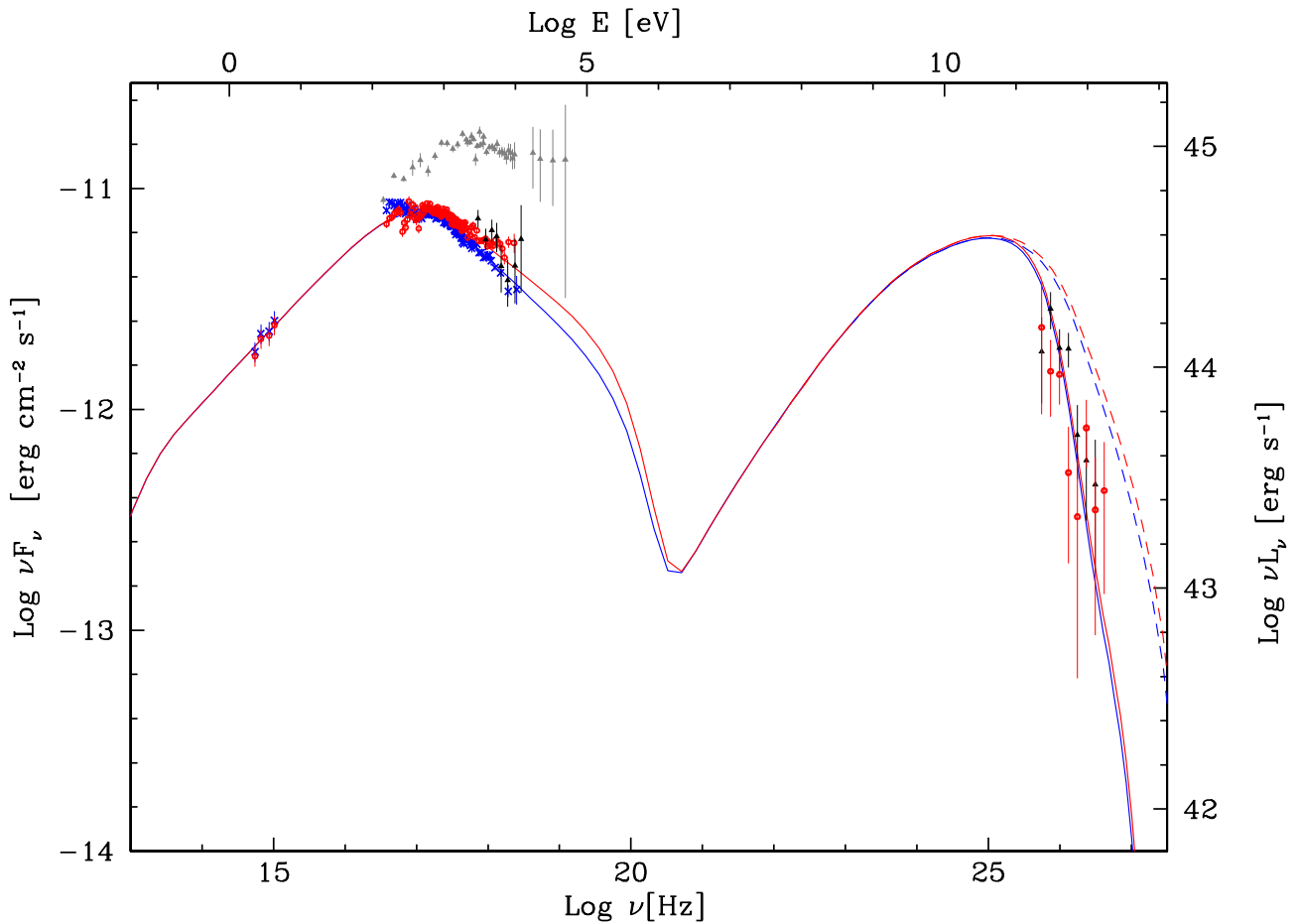}\\
      \caption{SED of H\,2356$-$309 in different epochs. The
        black triangles represent the SED from the 2004 RXTE campaign.  
	The closed-blue and open-red circles represent
	the SED from the 2005 XMM-Newton observations on MJD 53534 and 53536, respectively.
	The light-gray triangles correspond to the BeppoSAX X-ray data during the high state
	observed in June 1998 \cite{SAX_data}.
	In the VHE band, the black triangles and red circles are the
	time-average spectra observed in 2004 and 2005, respectively, and are not
	corrected for the effects of EBL absorption. 
	The curves are fits to the respective SEDs of a single-zone homogeneous
	SSC model described in the text.  The solid and dashed curves represent
	the same model with and without, respectively, the EBL effects included (\cite{Franceschini} 2008).
	The SSC model fit to the 2004 RXTE campaign data is identical to the
	fit to the MJD 53536 data (red curve). H\,2356-309 
	does not appear in the Fermi-LAT bright-AGN list \cite{LAT_AGN}, 
	and only poorly-constrained, non-simultaneous spectral data exist
	in the first Fermi-LAT catalog \cite{LAT_1FGL}. 
	The NRT (11 cm) measurement
	falls well below the minimum flux density of the figure and is not
        shown for the sake of clarity.  For the luminosity axis, the measured
	redshift of z=0.165 is used
	with H = 70 km\,s$^{-1}$\,Mpc$^{-1}$, $\Omega_M=0.3$. }
        \label{H2356_SEDs}
  \end{figure*}

\begin{figure}
  \centering
     \includegraphics[width=0.33\textwidth, angle = 270]{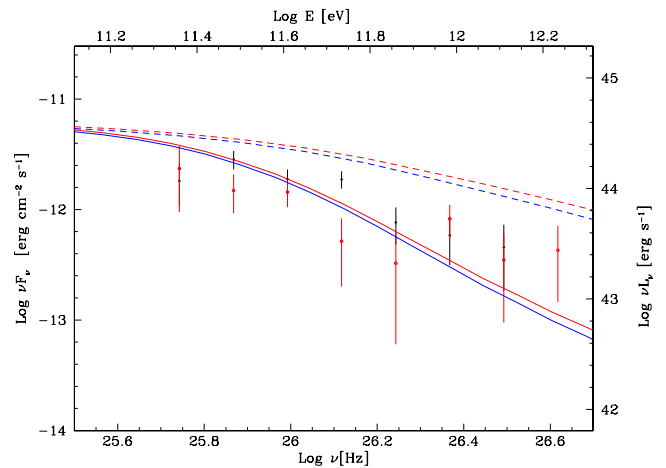}
     \caption{SED of H\,2356-309 zoomed in on the VHE band.  The data and model curves
are identical to those in Figure~\ref{H2356_SEDs}.}
        \label{VHE_zoom}
  \end{figure}

The broad-band SED of H\,2356$-$309 is shown in 
Figure~\ref{H2356_SEDs} and Figure~\ref{VHE_zoom} for the RXTE campaign in 2004 
and for each of the two XMM-Newton observations in 2005.  All of these
SEDs have the same shape as most VHE HBL, indicating a double-peaked structure
with a roughly similar energy output in both the keV and VHE range.
It should be noted that the optical, X-ray and VHE measurements
from each of the three campaigns (RXTE, XMM-Newton on MJD 53534 and
XMM-Newton on MJD 53536) are very similar. The measured X-ray flux 
from 2 to 10 keV differs by only $\pm15$\% from the mean value, 
and the X-ray spectral slopes are consistent above 2 keV. 
The annual VHE fluxes differ by less than  $\pm30\%$ from the mean 
and the annual photon indexes are constant in time.
As the VHE and X-ray spectral slopes are all consistent, 
and only modest changes in the VHE and X-ray fluxes are observed,
it is expected that modeling of each of the three SEDs
will result in similar physical parameters.

The aforementioned characterization of these three SEDs is performed
using a simple, time-independent SSC model (\cite{Katarzynski}).
In this SSC scenario, the optical through VHE emission is
generated by a spherical emission region of radius $R$ 
relativistically propagating with a Doppler factor $\delta$ with respect to the observer
through a homogeneous magnetic field $B$. The emission region is filled with
relativistic electrons that generate the lower-energy peak at X-ray frequencies
via synchrotron radiation and the higher-energy peak at VHE frequencies
through inverse-Compton scattering of the synchrotron photons off
the same electrons.  The high-energy electron distribution 
between Lorentz factors $\gamma_{\rm min}$ and $\gamma_{\rm max}$ is
assumed to have a broken power-law shape with normalization $K$ and break $\gamma_{\rm b}$.
The energy index of the electrons is $n_1$ below $\gamma_{\rm b}$, and $n_2$ above $\gamma_{\rm b}$.  
A low-energy cutoff in the electron distribution of $\gamma_{\rm min}=1000$, is
chosen to prevent an increase of the inverse-Compton emission in the MeV-GeV domain,
that is not compatible with the non-contemporaneous upper limit on the blazar's MeV-GeV flux
derived from the first 5.5 months of Fermi-LAT data \cite{LAT_TeVAGN}.
The majority of the radio emission is assumed to be produced
from regions further out in the jet from the core.  Therefore the NRT radio measurements
are not used in the following SED modeling of the blazar, but can be considered
as upper limits for the blazar's radio flux.

As H\,2356$-$309 has a redshift of z=0.165, its observed
VHE spectrum is strongly affected by the absorption of VHE photons on the
EBL (see, e.g., the review of \cite{felix}).  These interactions 
($\gamma_{_{\mathrm{VHE}}}$\,$\gamma_{_{\mathrm{EBL}}}$\,$\rightarrow$\,$e^{+}$\,$e^{-}$; \cite{EBL_abs})
along the line-of-sight create an energy-dependent opacity $\tau(z,E)$ that 
is imprinted on the observed VHE spectrum 
($F_{\mathrm{obs}}(E) = F_{\mathrm{int}}(E)$\,e$^{-\tau(z,E)}$).
To remove this absorption in the SED modeling
the EBL density of \cite{Franceschini} (2008) is used to 
calculate the optical depths relevant for these observations of
H\,2356$-$309.  Fitting the de-absorbed time-averaged VHE spectrum to a power-law function
(dN/dE $\sim$ $E^{-\Gamma_{\rm int}}$)
yields a relatively hard intrinsic photon index of $\Gamma_{\rm int} = 1.97 \pm 0.15$.
Similar values of $\Gamma_{\rm int}$ are found when fitting
the de-absorbed data from the 2004 and 2005 VHE spectra used in the SSC modeling.
It should be noted that the approximate slope of each 
of the de-absorbed SSC curves (dashed lines in Figures \ref{H2356_SEDs} and \ref{VHE_zoom}) 
is slightly steeper than the power-law fit to only the HESS EBL-de-absorbed data.  
This is due to the inclusion of all multi-wavelength data in the curved SSC fits, 
whereas the power-law function is fit to only the HESS data.

The parameters of the SSC fit to the SED from each of the XMM-Newton campaigns 
are shown in Table~\ref{SSC_results}.  For the modeling of the RXTE campaign, 
the XMM-Newton optical spectrum from MJD 53536 is used instead of the average 
ROTSE-IIIc flux from the 2004 season, since they are consistent and 
the former more strongly constrains the fit.  In addition, the ROSTE-IIIc R-band
measurement is modeled from white light, whereas the XMM OM data are from filtered
light and hence are more appropriate for SED modeling. The data from
the RXTE campaign are well described by the parameters
of the SSC fit to the data from the second XMM-Newton night (MJD 53536)
and modification of these parameters does not yield an improved $\chi^2$.
Indeed the fit parameters do not vary greatly between the three campaigns,
as expected from the similarity of the measured quantities. 
This may suggest that a steady state of the blazar was observed, and that this
state can be explained by an SSC scenario.  The slight hardening of 
the XMM-Newton spectrum between MJD 53534 and 53536 is easily accounted 
for by only a minor variation in the high-energy index of the electron distribution.

It should also be noted that the parameters of the SSC fits indicate a cooling time scale 
of $\sim$$1.5 \times 10^5$ s ($\sim$4 h), which is comparable to 
the source crossing time ($t_{esc} = R / c = 2.5 \times  10^5$ s), 
but is considerably shorter than the monthly time scale of the weak variations 
observed in the VHE band.  However, VHE flux variations 
on time scales comparable to the cooling time (e.g. daily), and of approximately
the same magnitude, cannot be ruled out due to the low average VHE flux.

  \begin{table}
    \begin{minipage}[t]{1.0\columnwidth}
     \caption{Results of SSC fits to the SED of H\,2356$-$309.}
        \label{SSC_results}
       \centering
        \begin{tabular}{c c c c c}
           \hline\hline
           \noalign{\smallskip}
		& XMM$_1$\footnote{The XMM$_1$ and XMM$_2$ results refer to the fits to the data
from MJD 53534 and 53536, respectively.  Using the XMM-Newton optical
spectrum from MJD 53536 instead of the average ROTSE-IIIc flux from 2004 for the RXTE campaign
yields the same SSC fit parameters as shown in XMM$_2$.}  & XMM$_2$$^a$ & RXTE$^{\dagger}$ & AH06\footnote{This is the result of the original SSC modeling of
the RXTE campaign SED of H\,2356$-$309 published in \cite{HESS_MWLI}.  The RXTE$^{\dagger}$ results use the same multi-wavelength
data as AH06 along with an improved calibration of the VHE data. }\\
           \noalign{\smallskip}
           \hline
           \noalign{\smallskip}
	$\delta$ & 18 & 18 & 18 & 18\\
	$B$ [G] & 0.16 & 0.16 & 0.16 & 0.16\\
	$R$ [cm] & $7.5 \times 10^{15}$ &  $7.5 \times 10^{15}$ &  $7.0 \times 10^{15}$ & $3.4 \times 10^{15}$\\
	$K$ [cm$^{-3}$] & $8.0 \times 10^{4}$ &  $8.0 \times 10^{4}$ & $2.0 \times 10^{4}$ & $1.2 \times 10^{4}$\\
	$\gamma_{\rm min}$ &  $1.0 \times 10^{3}$ & $1.0 \times 10^{3}$ & $1.0 \times 10^{3}$ & $1.0 \times 10^{3}$\\
	$\gamma_{\rm b}$ & $1.0 \times 10^{5}$ &  $1.0 \times 10^{5}$ & $2.5 \times 10^{5}$ & $2.5 \times 10^{5}$\\
	$\gamma_{\rm max}$ & $3.0 \times 10^{6}$ & $3.0 \times 10^{6}$ & $3.0 \times 10^{6}$ & $3.0 \times 10^{6}$\\
	$n_1$ & 2.3 & 2.3 & 2.0 & 2.0 \\
	$n_2$ & 3.6 & 3.5 & 3.4 & 4.0 \\
           \noalign{\smallskip}
           \hline
      \end{tabular}
    \end{minipage}
  \end{table}

The comparison of the results of the 
fits to the RXTE campaign in 2004  (see XMM$_2$ in Table~\ref{SSC_results})
to the previously published results (\cite{HESS_MWLI}) shows some
differences.  These differences are not primarily due to the renormalization ($\sim$50\%
increase) of HESS flux from this epoch.  Rather they are
largely due to the improved knowledge of the optical spectrum of the blazar
from the XMM-Newton OM results.  Using the ROTSE-IIIc flux instead
of the XMM-Newton OM spectrum results in a different set 
of fit parameters  (see RXTE$^{\dagger}$ in Table~\ref{SSC_results}),
very similar to the previous modeling (see AH06 in Table~\ref{SSC_results}).  
However, this fit (RXTE$^{\dagger}$) falls significantly below the measured XMM-Newton optical spectrum.  
As the optical portion of the SED is not expected to change dramatically,
the characterization presented here (XMM$_2$) should be used
as a reference for future studies.

\section{Conclusion}

Long term HESS observations of H\,2356$-$309 have clearly confirmed the initial
VHE detection from 2004.  The source has a relatively low VHE flux (1.6\% Crab)
that is weakly variable (factor of $\sim$2) on time scales of at least months.  Although the flux is
variable, no changes in the VHE spectral slope ($\Gamma\approx3.1$) are found.
After accounting for the effects of the absorption of VHE photons on the extragalactic background
light, the intrinsic VHE spectrum of H\,2356$-$309 is found to be hard ($\Gamma_{\rm int}\approx2.0$).  
Interestingly, the X-ray spectrum is
also very hard, with the synchrotron peak located at $\sim$1 keV, despite the source
being in a historically low state in X-rays.  
While the synchrotron peak of H\,2356$-$309 
is located at higher than typical energies and 
the X-ray and VHE spectra are comparatively hard, 
the SED can be reasonably modeled using a simple one-zone SSC scenario. 
This could be expected since the luminosity of both SED peaks are similar, 
and their separation in frequency is not extreme.

\begin{acknowledgements}
The support of the Namibian authorities and of the University of Namibia
in facilitating the construction and operation of H.E.S.S. is gratefully
acknowledged, as is the support by the German Ministry for Education and
Research (BMBF), the Max Planck Society, the French Ministry for Research,
the CNRS-IN2P3 and the Astroparticle Interdisciplinary Programme of the
CNRS, the U.K. Science and Technology Facilities Council (STFC),
the IPNP of the Charles University, the Polish Ministry of Science and 
Higher Education, the South African Department of
Science and Technology and National Research Foundation, and by the
University of Namibia. We appreciate the excellent work of the technical
support staff in Berlin, Durham, Hamburg, Heidelberg, Palaiseau, Paris,
Saclay, and in Namibia in the construction and operation of the
equipment.
\end{acknowledgements}


\begin{thebibliography}{}

\bibitem[(Abdo et al. 2009)]{LAT_AGN}
	Abdo, A. A., Ackermann, M., Ajello, M., et al. 2009, ApJ, 700, 597

\bibitem[(Abdo et al. 2009)]{LAT_TeVAGN}
	Abdo, A. A., Ackermann, M., Ajello, M., et al. 2009, ApJ, 707, 1310

\bibitem[(Abdo et al. 2009)]{LAT_1FGL}
	Abdo, A. A., Ackermann, M., Ajello, M., et al. 2010, ApJS, submitted, [arXiv:astro-ph/1002.2280]

\bibitem[Aharonian 2001]{felix}
       Aharonian, F. 2001, Proceedings 27th ICRC (Hamburg), 
       Invited, Rapporteur, and Highlight Papers, 250

\bibitem[Aharonian et al. (2004)]{calib_paper}
       Aharonian, F., Akhperjanian, A.G., Aye ,K.-M., et al. 
	(HESS Collaboration) 2004, Astroparticle Physics, 22, 109

\bibitem[Aharonian et al. 2005]{HESS_421}
       Aharonian, F., Akhperjanian, A.G., Aye, K.-M., et al. 
	(HESS Collaboration) 2005, A\&A, 437, 95

\bibitem[Aharonian et al. 2006a]{HESS_discovery}
       Aharonian, F., Akhperjanian, A.G., Bazer-Bachi ,A.R., et al. 
	(HESS Collaboration) 2006a, Nature, 440, 1018

\bibitem[Aharonian et al. 2006b]{HESS_MWLI}
       Aharonian, F., Akhperjanian, A.G., Bazer-Bachi, A.R., et al. 
	(HESS Collaboration) 2006b, A\&A, 455, 461

\bibitem[Aharonian et al. 2006c]{HESS_crab}
       Aharonian, F., Akhperjanian, A.G., Bazer-Bachi, A.R., et al. 
	(HESS Collaboration) 2006c, A\&A, 457, 899

\bibitem[Aharonian et al. 2007a]{2155_flare}
       Aharonian, F., Akhperjanian, A.G., Bazer-Bachi, A.R., et al. 
	(HESS Collaboration) 2007a, ApJ, 664, L71 

\bibitem[Benbow 2005]{std_analysis} 
	Benbow, W. 2005, Proceedings of Towards a Network of 
       Atmospheric Cherenkov Detectors VII (Palaiseau), 163

\bibitem[(Berge et al. 2007)]{bgmodel_paper} 
	Berge, D., Funk, S. \& Hinton, J. 2007, 
	A\&A, 466, 1219

\bibitem[Cardelli et al. (1989)]{reddening}
	Cardelli, J.A., Clayton, G.C., \& Mathis, J.S. 1989, ApJ, 345, 245

\bibitem[Cheung et al. 2003]{optical_res2}
	Cheung, C.C., Urry, C.M., Scarpa, R. \& Giavalisco, M. 2003, ApJ, 599, 155

\bibitem[(Costamante et al. 2001)]{SAX_data}
       Costamante, L., Ghisellini, G., Giommi, P., et al. 2001, A\&A, 371, 512

\bibitem[Costamante \& Ghisellini (2002)]{Costamante_AGN}
       Costamante, L. \& Ghisellini, G. 2002, A\&A, 384, 56

\bibitem[Falomo 1991]{redshift}
	Falomo, R. 1991, AJ, 101, 821

\bibitem[Fazio et al. 2004]{fazio} 
       Fazio, G.G., Ashby, M.L.N., Barmby, P., et al. 2004, ApJS, 154, 39

\bibitem[(Feldman \& Cousins 1998)]{UL_tech} 
	Feldman, G.J. \& Cousins, R.D. 1998,
       Phys Rev D, 57, 3873

\bibitem[(Forman et al. 1978)]{UHURU_det}
	Forman, W., Jones, C., Cominsky, L., et al. 1978, ApJS, 38, 357

\bibitem[Franceschini]{Franceschini}
	Franceschini, A., Rodighiero, G., \& Vaccari, M. 2008, A\&A, 487, 837 

\bibitem[Fukugita et al. (1995)]{spectral_template}
	Fukugita, M., Shimasaku, K. \& Ichikawa, T. 1995, PASP, 107, 945

\bibitem[(Giommi et al. 2005)]{Sedentary}
	Giommi, P., Piranomonte, S., Perri, M. \& Padovani, P. 2005, A\&A, 434, 385 

\bibitem[Gould \& Schr\'eder 1967]{EBL_abs}
       Gould, R.J. \& Schr\'eder, G.P. 1967, Physical Review, 155, 1408

\bibitem[(Hauser et al. 2004)]{ATOM} 
	Hauser, M., M\"ollenhoff, C., P\"uhlhofer, G., et al. 2004, Astronomische Nachrichten, 325, 659

\bibitem[Kalberla et al. 2005]{gal_abs}
	Kalberla, P.M.W., Burton, W.B., Hartmann, D., et al. 2005, A\&A, 440, 775

\bibitem[Katarzy\'nski et al. 2001]{Katarzynski}
	Katarzy\'nski, K., Sol, H. \& Kus, A. 2001, A\&A, 367, 809

\bibitem[Li \& Ma (1983)]{lima} Li, T. \& Ma, Y. 1983, ApJ, 272, 317

\bibitem[O'Donnell (1994)]{reddening_update}
	O'Donnell, J.E. 1994, ApJ, 422, 158

\bibitem[Primack et al. (2005)]{Primack_05}
      Primack, J.R., Bullock, J.S. \& Somerville, R.S. 2005, 
      AIP Conference Proceedings, 745, 23

\bibitem[Scarpa et al. 2000]{optical_res}
	Scarpa, R., Urry, C.M., Padovani, P., Calzetti, D. \& O'Dowd, M. 2000, ApJ, 544, 258

\bibitem[Theureau et al. 2007]{Nancay1}
	Theureau, G., Hanski, M. O., Coudreau, N., Hallet, N. \& Martin, J.-M. 2007, A\&A, 465, 71

\bibitem[Urry et al. (2000)]{Urry_mag}	
	Urry, C.M., Scarpa, R., O'Dowd, M., et al. 2000, ApJ, 532, 816

\bibitem[Verner et al. (1996)]{verner}
	Verner, D.A., Ferland, G.J., Korista, K.T. \& Yakovlev, D.G. 1996, ApJ, 465, 487

\bibitem[Wilms et al. (2000)]{wilms}
	Wilms, J., Allen, A., \& McCray, R. 2000, ApJ, 542, 914

\bibitem[(Wood et al. 1984)]{HEAO_det}
	Wood, K.S., Meekins, J.F., Yentis, D.J., et al. 1984, ApJS, 56, 507

\bibitem[(Zacharias et al. 2005)]{NOMAD}
	Zacharias, N., Monet, D.G., Levine., S., et al. 2005, VizieR Online Data Catalog, 1297, 0

\end{thebibliography}
\end{document}